\documentclass[mathleft,twocolumn]{an}
\usepackage{graphicx}
\usepackage{times}

\begin{document}

\title{The Evolution of Cluster Dwarfs}

\author{Daniel Harsono\inst{1}\fnmsep\thanks{
  \email{dan.harsono@gmail.com}\newline}
\and Roberto De Propris \inst{2}
}
\titlerunning{The Evolution of Cluster Dwarfs}
\authorrunning{Daniel Harsono and Roberto De Propris}
\institute{
$^1$ Leiden Observatory, Leiden University, P.O. Box 9513, 2300RA Leiden, The Netherlands\\
$^2$ Cerro Tololo Inter-American Observatory, Casilla 603, La Serena, Chile}


\keywords{galaxies: luminosity function, mass function -- galaxies: dwarf -- galaxies: evolution -- galaxies: clusters}

\abstract{%
We summarize the results from analyzing six clusters of galaxies at $0.14 < z < 0.40$ observed with the Hubble Space Telescope Advanced Camera for Surveys. We derive deep composite luminosity functions in B, g, V, r, i and z down to absolute magnitude of $\sim$ -14 + 5 log \textit{h} mag. The luminosity functions are fitted by a single Schechter function with $M_{BgVriz}^{*} = -19.8,\ -20.9, \ -21.9, \ -22.0, \ -21.7,$ and $-22.3$ mag. and $\alpha \ \sim \ -1.3$ for all bands. The data suggests red sequence dominates the luminosity function down to $ \ge 6$ mag. below $L_*$, the dwarf spheroidals regime. Hence, at least at $z \sim 0.3$, the red sequence is well established and galaxies down to dwarf spheroidals are assembled within these clusters. We do not detect the faint-end upturn ($M > -16$) that is 
observed in lower redshift clusters. If this is real, the faint-end population has originated since $z=0.3$.}

\maketitle

\section{Introduction}

Clusters of galaxies are a unique environment to explore the formation and evolution of galaxies,
as they provide a volume-limited sample of galaxies at the highest density peaks as a function of
redshift, and therefore allows us to study a consistent population of object to cosmologically
significant lookback times. The cluster environment, however, has a strong impact on the star
formation history (Lewis et al. 2002, Gomez et al. 2003, Balogh et al. 2004) and morphology
(e.g., Dressler 1980). In order to understand the effects of environment, it repays to
observe the faintest cluster galaxies, dwarfs, which are particularly sensitive to interactions
with the cluster tidal field and other galaxies  (`harassment') and/or the intracluster gas (e.g., 
ram stripping). Observations of more distant clusters allows us to see this process, `as it happens' 
and understand the process of galaxy formation and evolution, while making due allowance for the high 
cluster densities.

The Luminosity Function (LF) of galaxies in clusters has long been used as a tool to probe galaxy
evolution, via its shape parameters $M^*$ and $\alpha$. The characteristic luminosity $M^*$ is 
sensitive to the evolution of massive galaxies, while $\alpha$ best describes the dwarf galaxy
population. It has now been established that massive galaxies ($M > M^*$) clearly form their
stellar populations and assemble their mass at high redshift (e.g., De Propris et al. 2007,
Muzzin et al. 2008). This is consistent with observations of `downsizing' in the general field
(Perez-Gonzalez et al. 2008). The evolution of dwarf galaxies is much more uncertain. One expects
the initial slope of the dwarf galaxy LF to reflect the steep slope that emerges from CDM halos
at the time of recombination, $\alpha \sim -2$, and that the slope subsequently flattens as
dwarf galaxies are destroyed in dense environments to fuel the formation of giant galaxies
(Khochfar et al. 2007). There is some evidence that the red sequence is truncated in more
distant clusters (De Lucia et al. 2007, Stott et al. 2007, Krick et al. 2008), but this does
not appear to be true for all clusters (Andreon 2006, Crawford et al. 2009) and may be due
to a selection effect against faint red galaxies in the blue bandpasses. Even if the red 
sequence is truncated, it does not imply that the slope actually grows less steep at high
redshift, which would be a severe falsification of hierarchical formation models. The dwarf
galaxies on the red sequence may lie in the blue cloud until their star formation is truncated.
In the field, the space density of red galaxies appears to grow by a factor of 2 since $z \sim 
1$, mostly at the faint end of the red sequence and via star formation suppression (Bell et al.
2006, Faber et al. 2007).

Related to this is the issue of the presence of a faint-end upturn in local cluster populations.
This is still a controversial finding, but it has since been confirmed by deep composite LFs of
clusters by Popesso et al. (2006) and Barkhouse et al. (2007). If the local upturn is indeed real,
it is interesting to see whether the upturn also exists at higher redshifts and whether the
upturn galaxies are on the red sequence or in the blue cloud. 

To this purpose we have started an investigation of deep luminosity functions and color-magnitude
diagrams for a sample of six clusters at $0.14 < z < 0.40$ with deep, multicolor HST ACS archival
data. The deep LFs will allow us to explore the evolution of dwarf galaxies, while the color
magnitude relations will be used to compute the scatter as a function of luminosity and derive
the LFs for red sequence and blue cloud galaxies, and thus set limits on the mechanism(s) by which
the faint end of the red sequence is constructed.

The following section summarizes the data, their reduction, analysis and photometry. The discussion of 
the results and interpretation can be found in Section 3. We adopt $\Omega_M = 0.27, ~ \Omega_{\Lambda} 
= 0.73$ and $H_0 = 100\mathrm{ \ km \ s^{-1} \ Mpc^{-1}}$ cosmology. All data are photometrically calibrated 
to the AB magnitude system.

\section{Observations and Data Analysis}

We analyzed images of six clusters (Abell 1413, 2218, 1689, 1703, MS1358+62 and Cl0024+17) at $0.14 < z < 0.40$ 
taken with Advanced Camera for Surveys (ACS) on board the Hubble Space Telescope (HST). The archival images
are taken in filters $B,g,V,r,i$ and $z$ with exposure times of 5--20ks in each color. We retrieved the
flat fielded exposures from the HST archive and processed them through {\tt Multidrizzle} (Koekemoer et
al. 2002), producing a single images with the gaps between chips interopolated and cosmic rays removed.
Figure 1 shows a false color image of one of our clusters (Abell 2218).

\begin{figure*}
\centering
\includegraphics[width=6in]{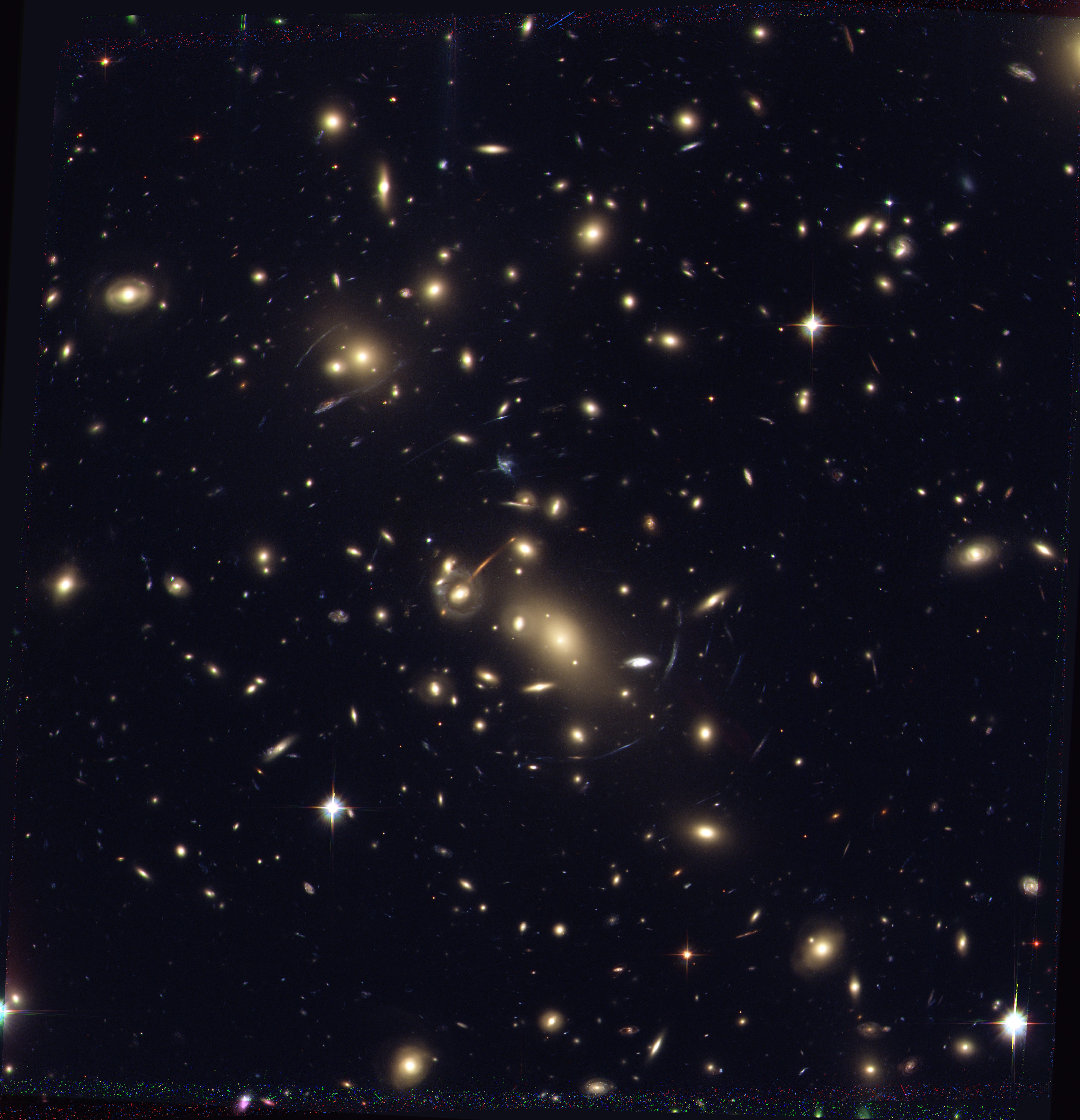}
\caption{A false color image (B=blue, V=green, z=red) for A2218 from ACS imaging}
\end{figure*}

Detection and photometry were carried out using SEXTRACTOR (Bertin \& Arnouts 1996). We used a minimum detection 
`aperture' of 7 connected pixels $2.0\ \sigma$ above the sky. We also calculated a mean central surface brigthness
from a small aperture (of area equivalent to the detection aperture) for star-galaxy separation. We experimented 
with the parameters to maximize the number of faint objects detections and minimize the number of spurious objects. 
All detections were examined visually to eliminate noise spikes due to drizzling process, bleed trails from bright stars, 
arclets (especially !) and false detections. The magnitudes were calibrated to the HST AB system using the published 
zeropoints on the HST Instrument Web page. We also corrected the magnitudes to SDSS and Johnson-Cousins systems using 
the values tabulated in Holberg \& Bergeron (2006). 

We use plots in surface brightness and magnitude space to select a range of luminosities and central surface
brightnesses where we are reasonably complete, can separate stars and galaxies and do not suffer excessively
from incompleteness at low surface brightness. See Harsono \& De Propris (2009) for details.

We can only determine cluster memberships statistically, especially at the faint magnitudes. The `field' galaxies 
can be removed by observing cluster-less fields. We use the two GOODS fields (Giavalisco et al. 2004) as our 
reference fields, which have deeper multicolor data. We use the same SExtractor parameters on the GOODS images 
as the ones used for the clusters. We selected the galaxies in GOODS fields using the limiting magnitudes and
surface brightness limits we used for clusters.

The GOODS number counts were fitted with a quadratic fit to smooth out field to field variations in the counts. 
The reference counts were scaled to the areas surveyed for each cluster and these counts subtracted from the 
galaxy counts in each cluster field to arrive at an estimate of the luminosity distribution for galaxies in each 
cluster. Errors include contributions from Poisson shot noise in the cluster, fore/background galaxies in the cluster 
field and galaxies in the reference field, and clustering errors in the cluster and reference fields, calculated as 
per Huang et al. (1997) and added in quadrature.

The magnitudes were $k+e$ corrected to $z = 0$ for comparison with local data using a Bruzual \& Charlot (2003) 
model for a solar metallicity single stellar population formed at $z = 3$ with an e-folding time of 1 Gyr (appropriate
for the bright cluster ellipticals). We then calculated composite LFs for each band using the approach in Colless
(1989). The LF fits are shown in Figure 2 and are consistent with a single Schechter function: the bottom panels
show the associated error ellipse. Table 1 shows the best fit values.

\begin{table}[!b]
 
 \centering
 \caption{Best Schechter Function Fit for Composite LFs}
 \label{tbl:sfits} 

\begin{tabular}{cccc}
 \hline
 \hline
 Band & $ M_* $ & $\alpha$ & $\chi_{\nu}^{2}$ \\
 \hline
 B (F435W) & $-19.89 \pm 0.27$ & $-1.29 \pm 0.06$ & 0.59 \\
 g (F475W) & $-20.94 \pm 0.17$ & $-1.31 \pm 0.04$ & 0.95 \\
 V (F555W) & $-21.86 \pm 0.27$ & $-1.27 \pm 0.05$ & 1.08 \\
 r (F625W) & $-21.95 \pm 0.29$ & $-1.33 \pm 0.03$ & 0.43 \\
 i (F775W) & $-21.66 \pm 0.27$ & $-1.27 \pm 0.04$ & 0.37 \\
 z (F850LP) & $-22.26 \pm 0.30$ & $-1.45 \pm 0.02$ & 0.94 \\
 \hline
\end{tabular}

\end{table}

\begin{figure*}[!t]

 \centering
 \begin{tabular}{cc}
  \includegraphics[scale=0.4]{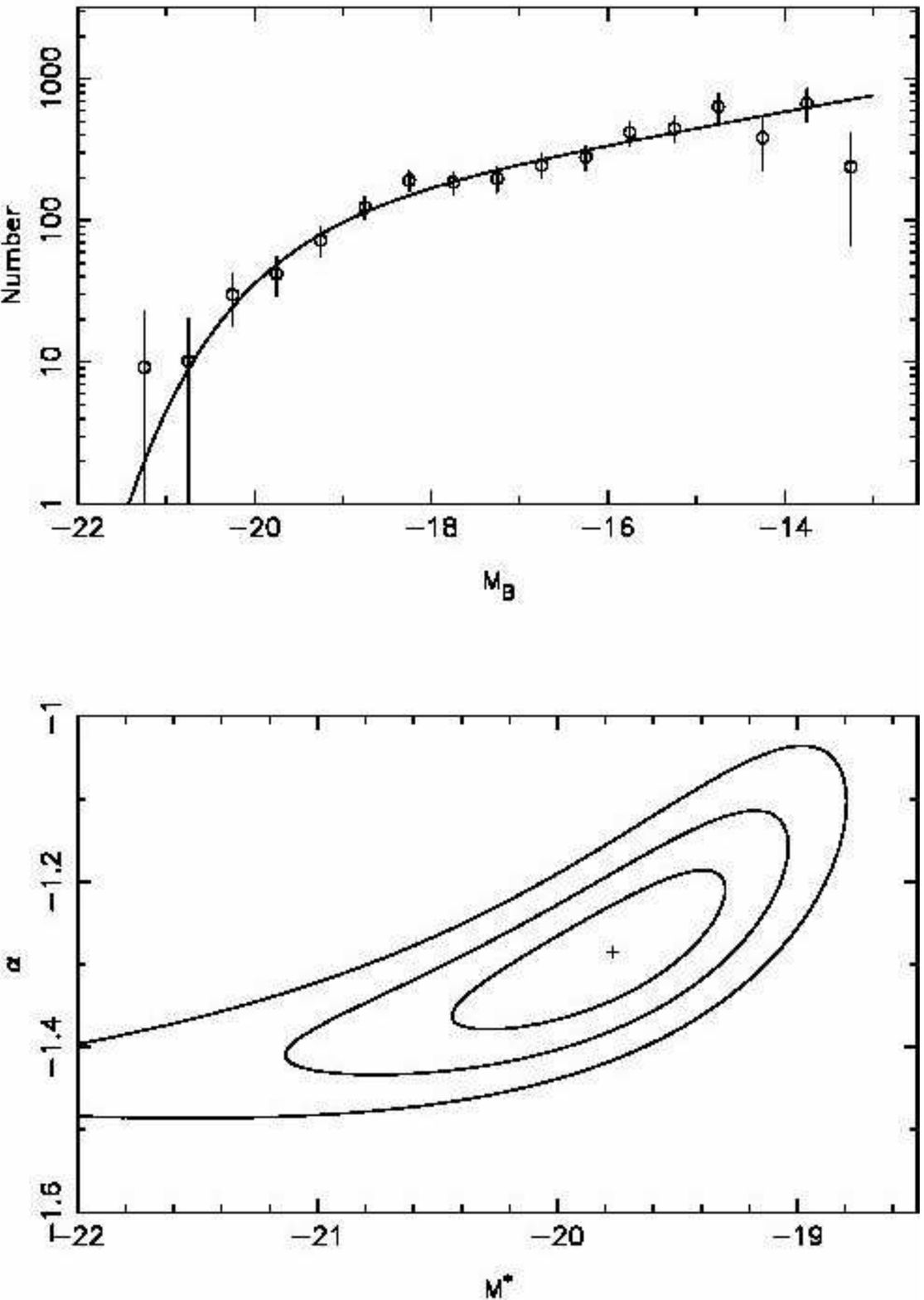} & \includegraphics[scale=0.4]{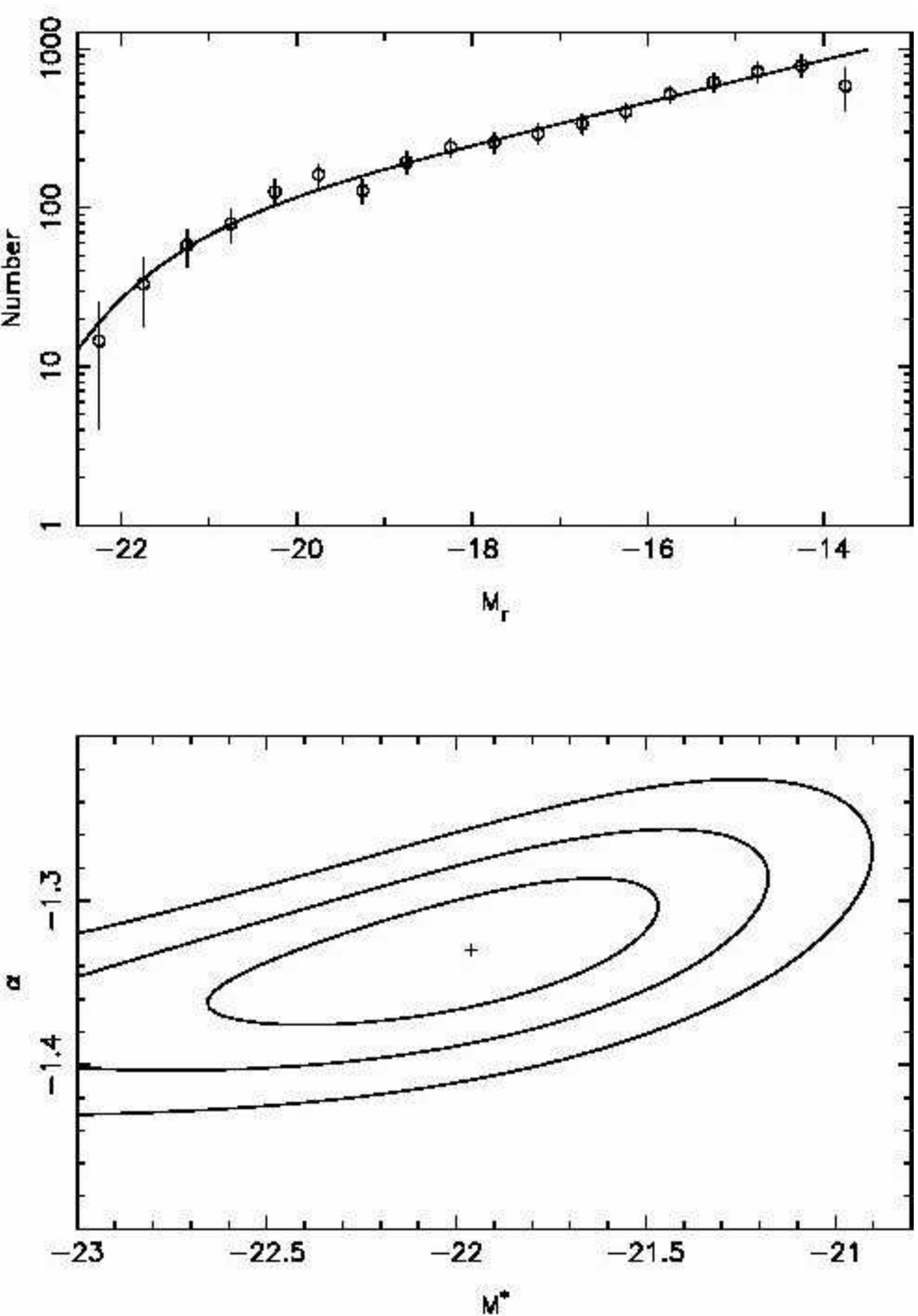} \\
  \includegraphics[scale=0.4]{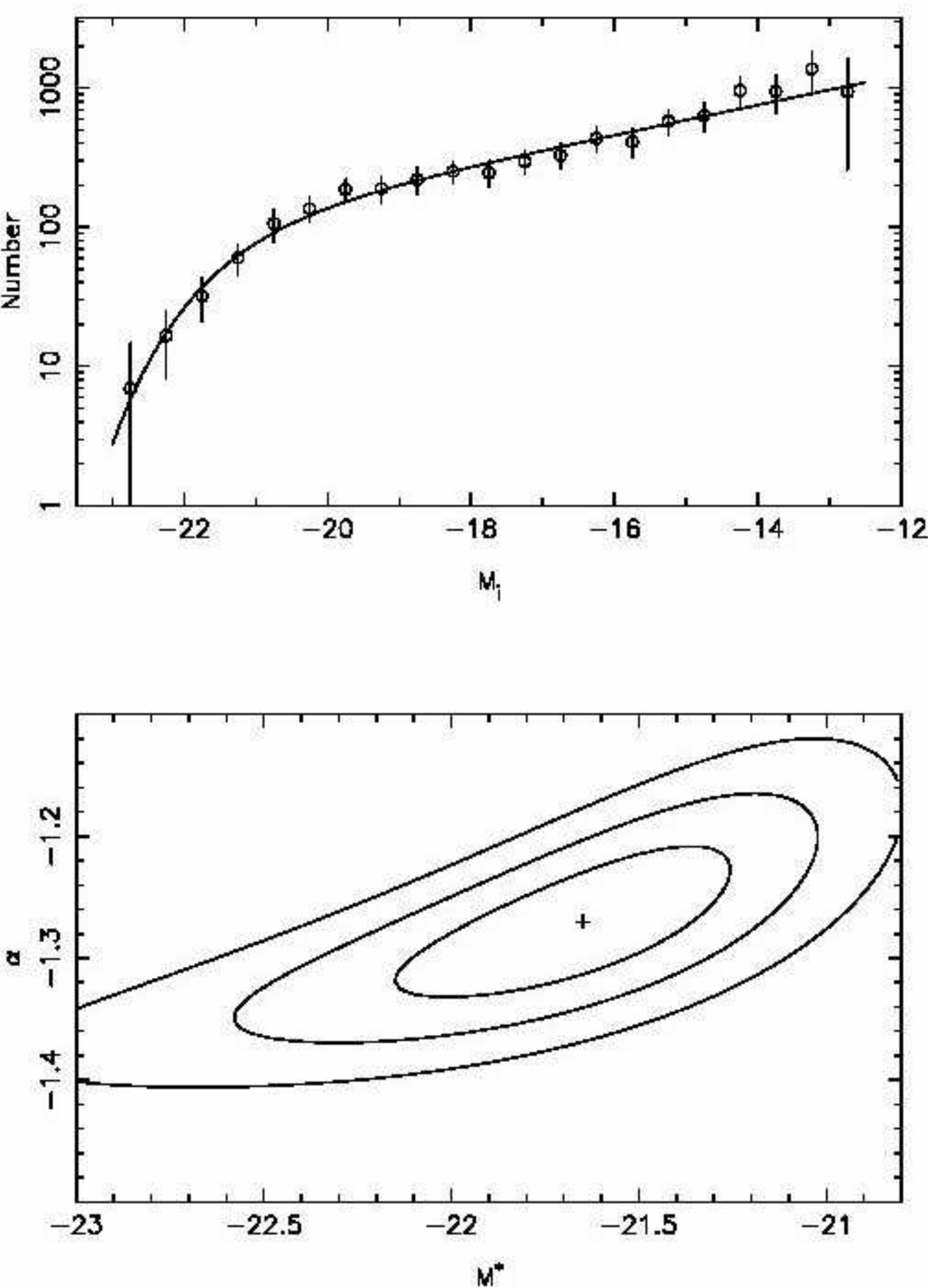} & \includegraphics[scale=0.4]{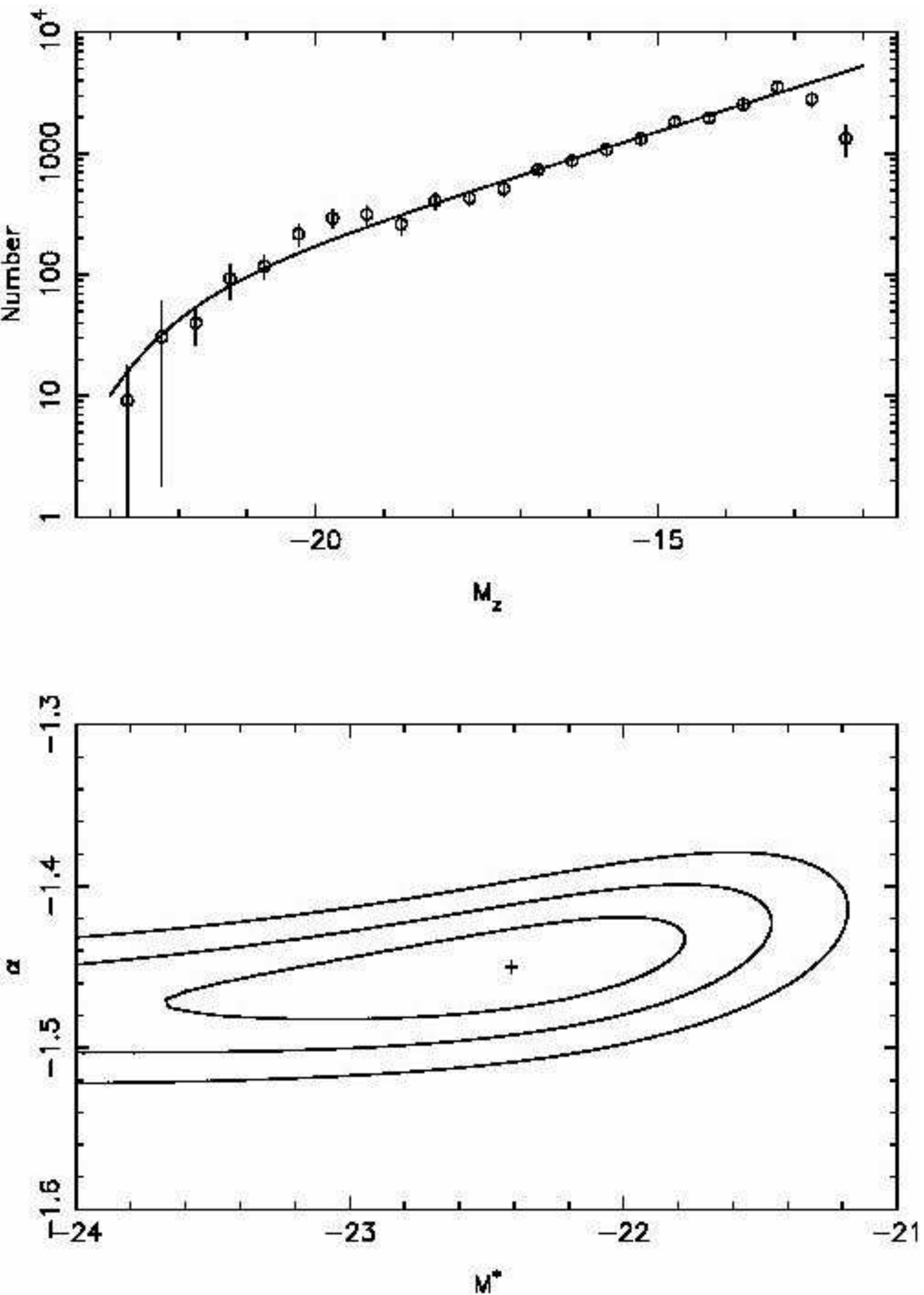} \\
 \end{tabular}
 
 \caption{Composite LFs, best fits, and error ellipses for B (top left), r (top right), i (bottom left) and z (bottom right).}
 \label{fig:lfs}
\end{figure*}

\section{Evolution of the LF}

Our data are consistent with a passively evolving bright-end of the LF, as has been already found
out by several studies. The main novel conclusion here regards the evolution of the faint end
slope, which is here measured to depths comparable to those reached in the nearby surveys by
De Propris et al. (2003) and Popesso et al. (2006). The data show that the LF in these clusters
has not evolved significantly since $z=0.3$. This suggests that the entire population of galaxies
in these cluster cores (the fields span the inner 1/3 of each cluster) was formed and assembled
at least at $z=0.3$ over a luminosity range spanning nearly 3 orders of magnitude.

We also note that the LF slope is the same in all bands. This argues that the cluster population
is dominated by galaxies on the red sequence at all luminosities and that there is no evidence for
truncation of the red sequence down to $M_z=-14$ in these clusters at $z=0.3$ and that the entire
red population must have been completely in place by this epoch at least.

We do {\it not} detect any sign of an upturn in the LF at faint magnitudes. If this upturn is truly
present in local clusters, the data suggest that it may have originated recently, by infall of a
very faint field population (Wilson et al. 1997).

We can explore this further by looking at the color-magnitude relation in population sensitive colors.
A full analysis of this is deferred to a future paper (Harsono, De Propris \& Andreon, in preparation).
Here we show (Figure 3) $g-r$ (a color that resembles $U-V$ for the redshifts of our clusters) vs. $z$ 
(a color that samples more closely the stellar mass of galaxies) for our clusters. We observe that all 
clusters obey a very tight and linear color magnitude relation, with little evidence of increasing scatter 
as a function of luminosity (cf., Andreon et al. 2006).

\begin{figure*}[!t]

 \centering
 \begin{tabular}{cc}
  \includegraphics[width=2.25in]{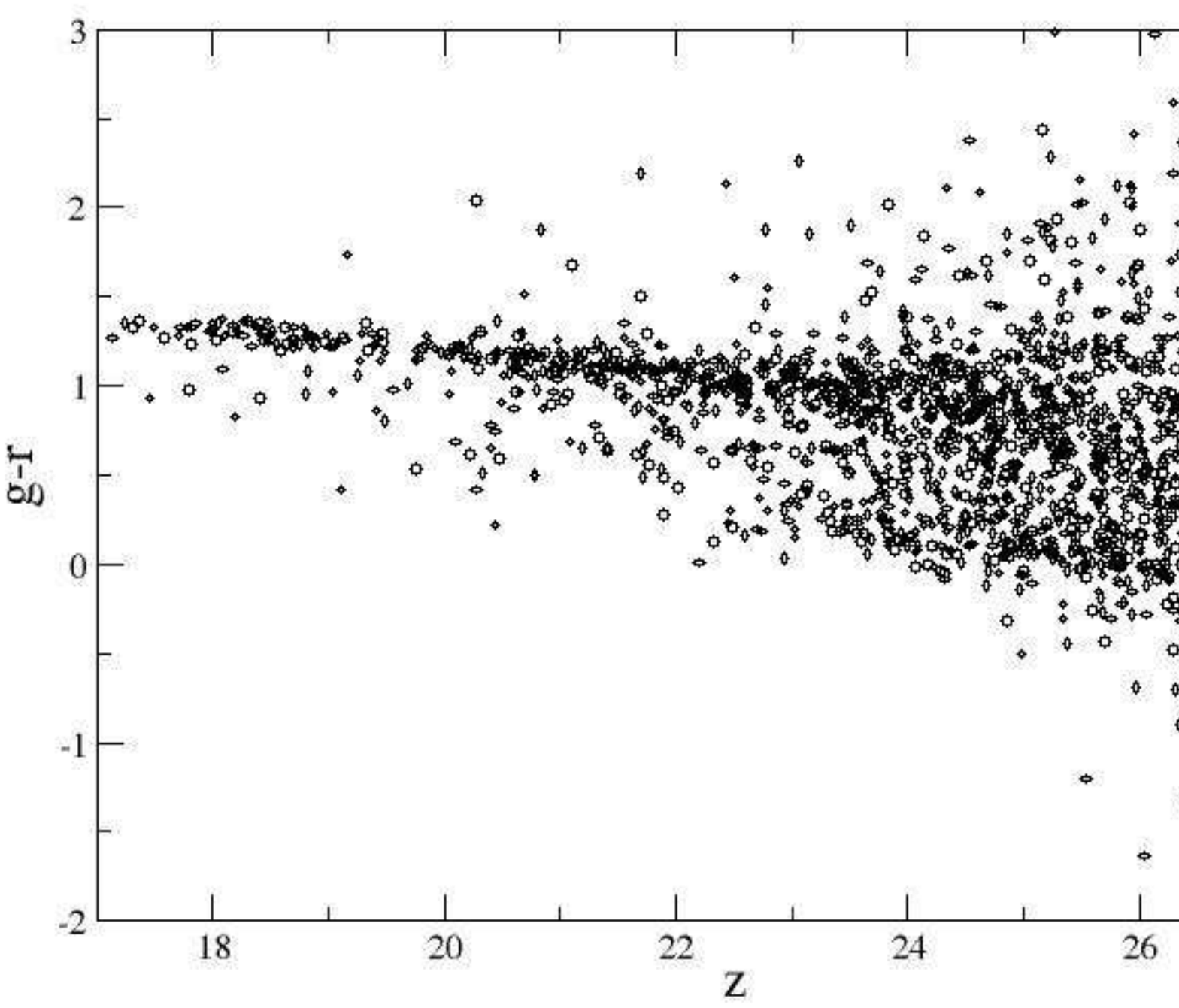} & \includegraphics[width=2.25in]{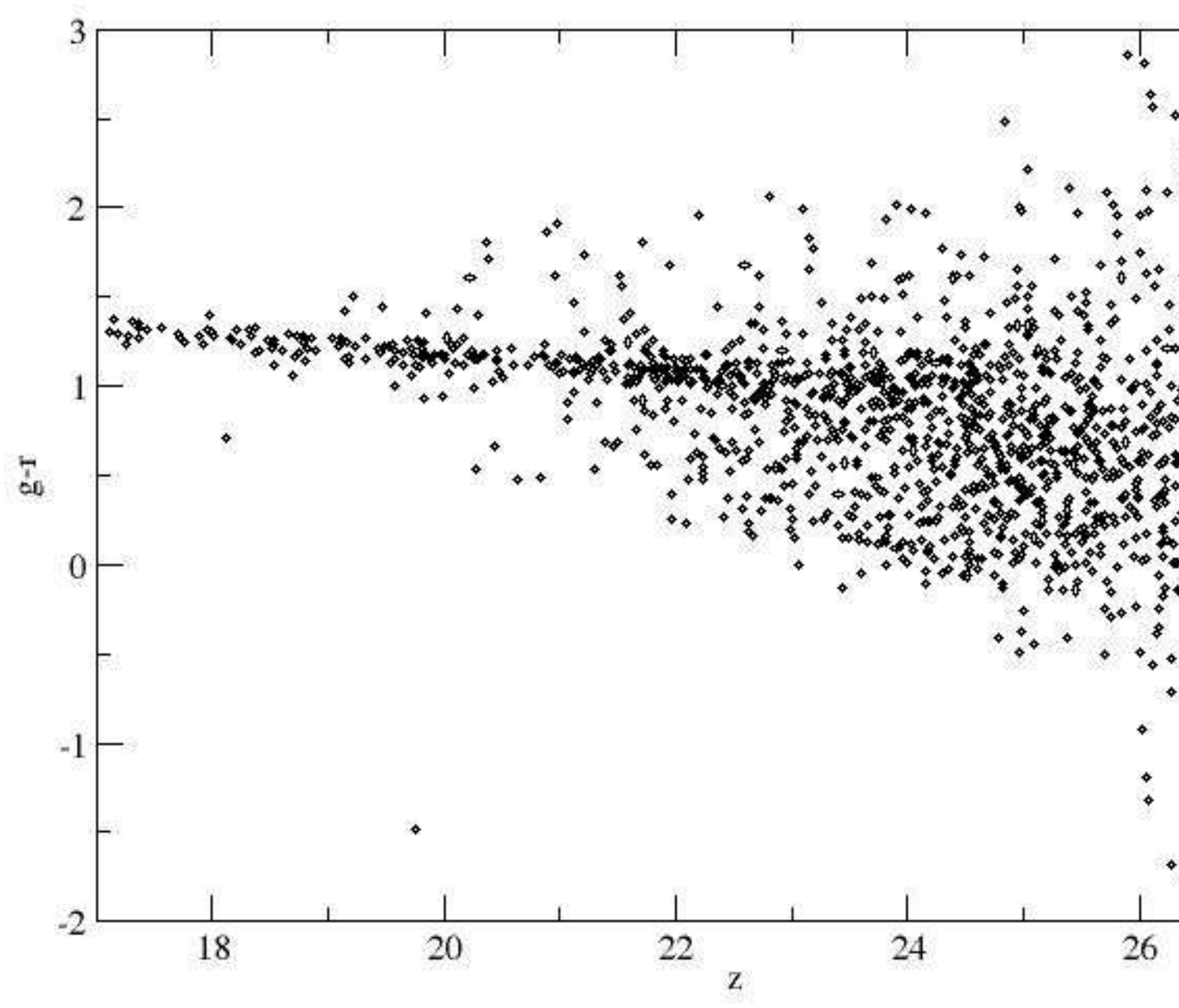} \\
  \includegraphics[width=2.25in]{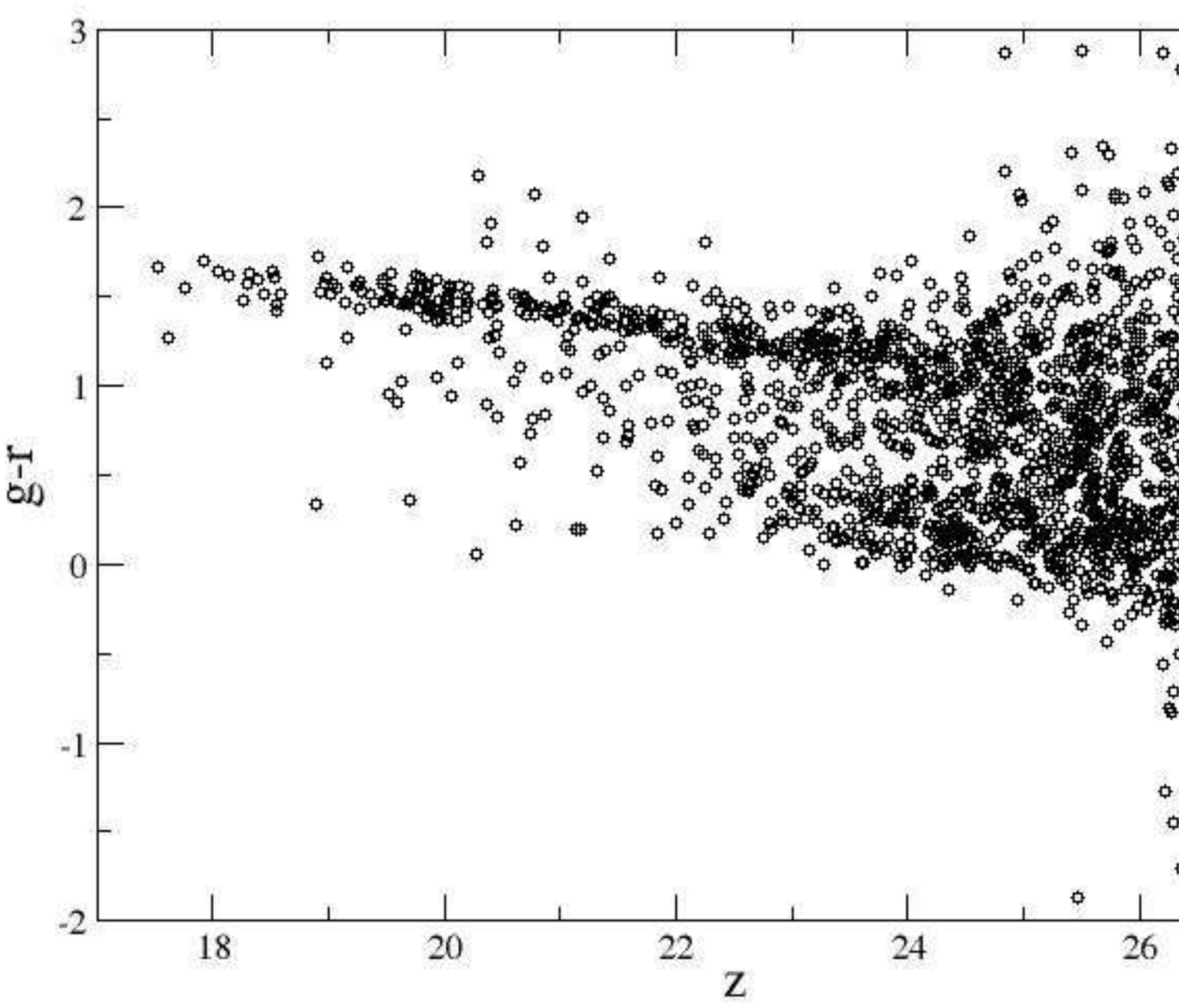} & \includegraphics[width=2.25in]{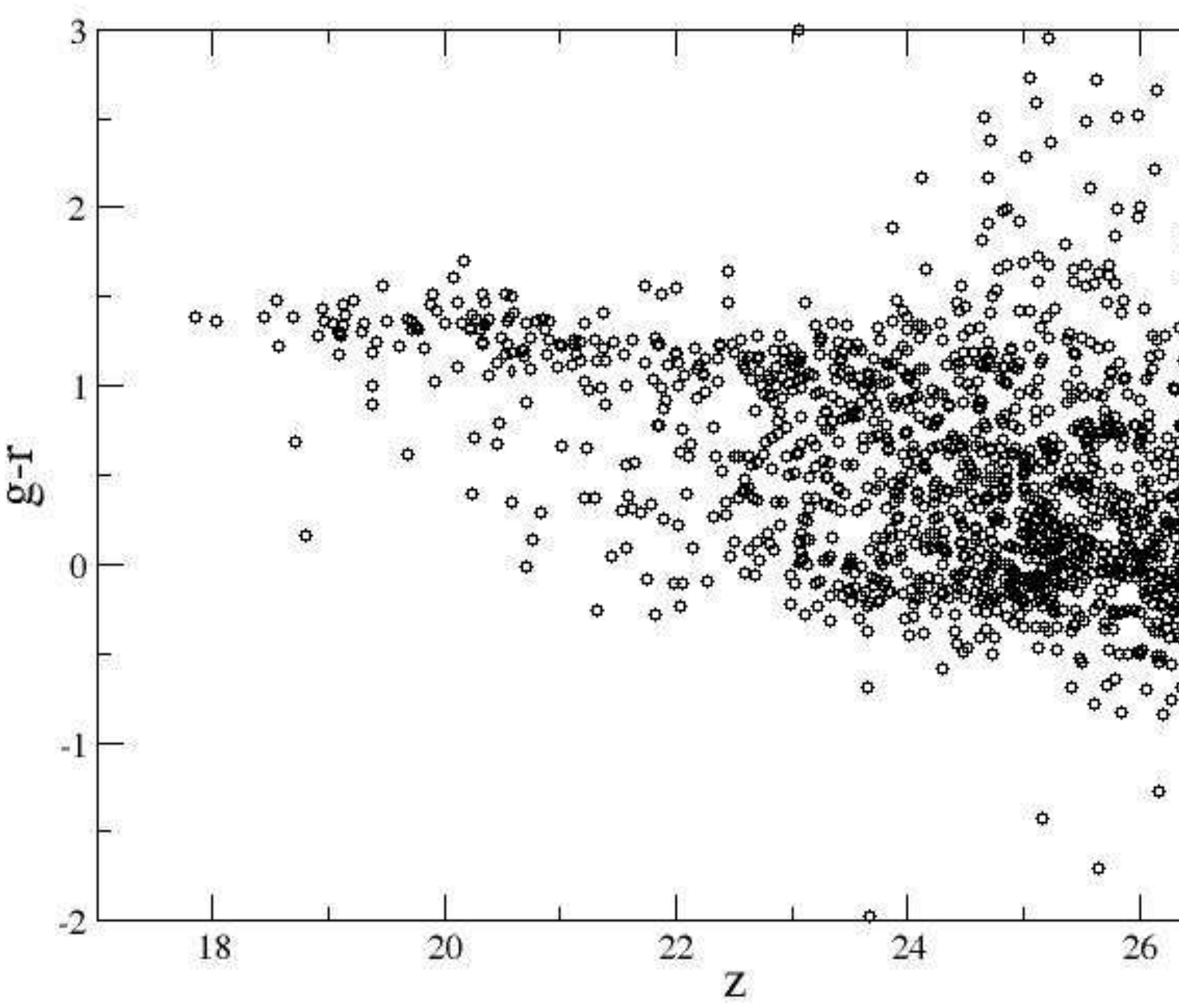} \\
  \includegraphics[width=2.25in]{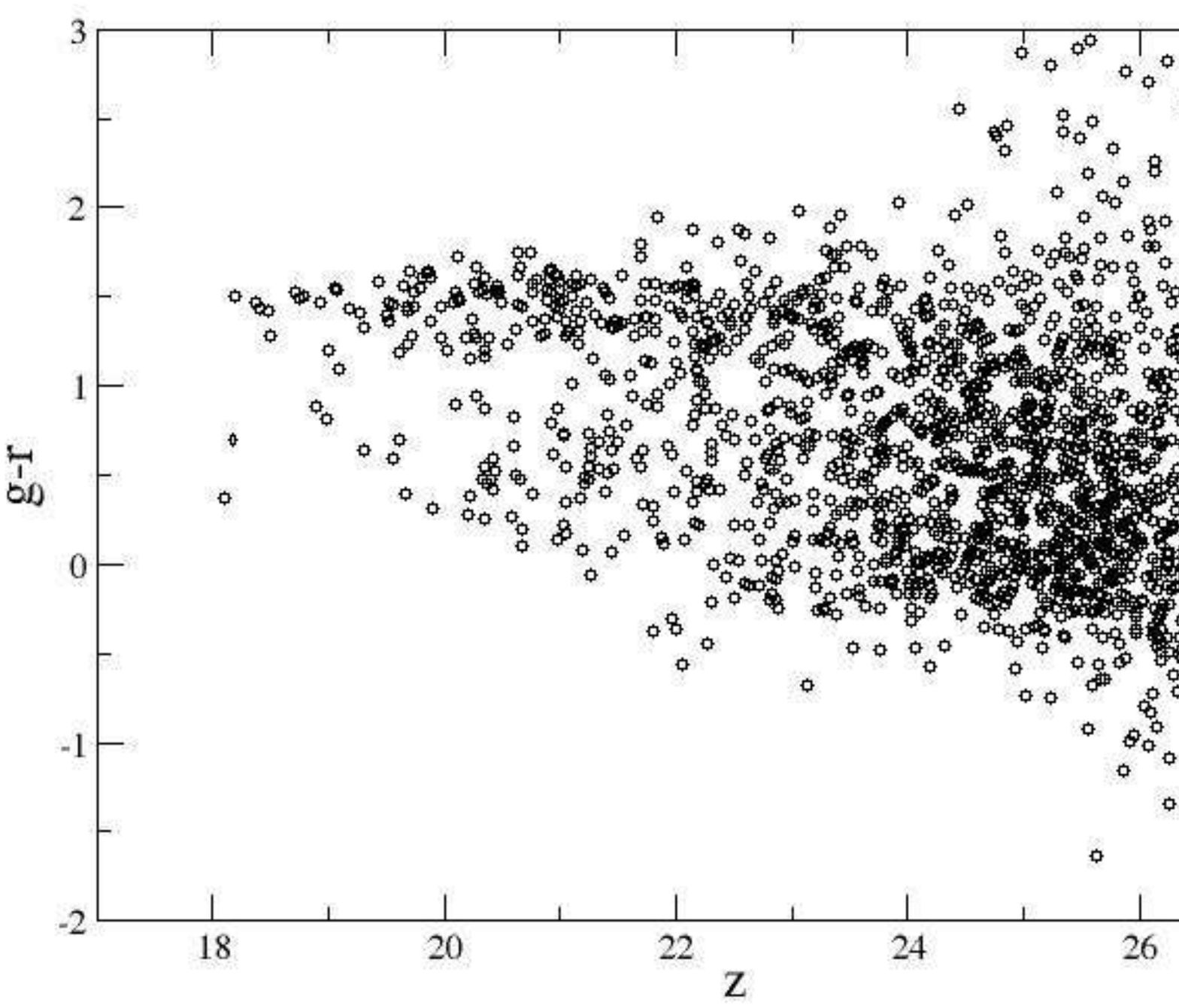} &  \\

 \end{tabular}
 
 \caption{The color magnitude diagrams (CMDs: g-r vs z) for 4 clusters: A1689 (top left), A2218 (top right), A1703 (middle
left), MS 1358.4+6245 (middle right) and Cl 0024.0+1652 (bottom left). We arranged CMDs such that the redshift increases
from left to right and top to bottom. }
 \label{fig:ind}
\end{figure*}

We are currently carrying out a rigorous Bayesian analysis to calculate the scatter of the color 
magnitude relation as a function of luminosity and the red sequence luminosity function. However,
if the above result is confirmed, this would suggest that the entire cluster population shared
a similar formation history, irrespective of galaxy mass, and that the red sequence was well in
place at faint luminosities even at $z > 1$ (cf., Bower et al. 1992).

\section{Future Work}

The limitations of the present study lie principally in the limited number of objects
studied and in the fact that the fields found in the archive correspond to the central
regions of very massive clusters and therefore constitute a biased sample: progenitor
like effects will be most evident in this sample. 

We wish to obtain HST data on the cores of poor clusters, chosen as being below the
knee of the X-ray luminosity function, with similar depth and band coverage, together
with parallel data on the cluster outskirts and, if feasible, a study of the cluster
outskirts of our rich cluster targets (analyzed here). This will allow us to study
how the LF evolves and depends on environment, and how the color magnitude relation
also evolves depending on environment.

An initial analysis of this can be carried out using a few existing archival fields.
This is part of our long term study of clusters at intermediate redshifts and will
be reported at a future time.

\newpage

\end{document}